\documentclass{osa-article-arxiv}

\journal{josaa-arxiv}
\articletype{Research Article}
\newcommand{\bd}{\begin{displaymath}}
\newcommand{\ed}{\end{displaymath}}
\newcommand{\be}{\begin{equation}}
\newcommand{\ee}{\end{equation}}
\newcommand{\bs}{\begin{subequations}}
\newcommand{\es}{\end{subequations}}
\newcommand{\ba}{\begin{eqnarray}}
\newcommand{\ea}{\end{eqnarray}}

\usepackage{lineno}
\begin{document}

\title{Flux trajectory analysis of Airy-type beams}

\author{\'Angel S. Sanz\authormark{*}}

\address{Department of Optics, Faculty of Physical Sciences,
Universidad Complutense de Madrid\\
Pza. Ciencias 1, Ciudad Universitaria 28040 Madrid, Spain}

\email{\authormark{*}a.s.sanz@fis.ucm.es} %% email address is required

% \homepage{http:...} %% author's URL, if desired

%%%%%%%%%%%%%%%%%%% abstract %%%%%%%%%%%%%%%%
%% [use \begin{abstract*}...\end{abstract*} if exempt from copyright]

\begin{abstract*}
Airy beams are solutions to the paraxial Helmholtz equation known for exhibiting
shape invariance along their self-accelerated propagation in free space.
These two properties are associated with the fact that they are not square integrable,
that is, they carry infinite energy.
To circumvent this drawback, families of so-called finite-energy Airy-type beams have been
proposed in the literature and, in some cases, also implemented in the laboratory.
Here an analysis of the propagation of this type of structured light beams is presented from
a flux trajectory perspective with the purpose to better understand the mechanisms that make
infinite and finite energy beams to exhibit different behaviors.
As it is shown, while the foremost part of the beam can be clearly and unambiguously 
associated with the well-known accelerating term, the rear part of the beam corresponds to
a nearly homogeneous distribution of flow trajectories, particularly for large propagation
distances.
This is shown to be related with an effective transfer of trajectories between adjacent lobes
(gradually, from the fore part of the beam to its rear part), which leads to smearing out the
transverse flow along the rear part of the beam.
This is sharp contrast with the situation found in ideal Airy beams, where trajectories
belonging to a given lobe of the intensity distribution remain the same all along the
propagation.
The analysis is supplemented with an also trajectory-based description of Young's
experiment performed with finite-energy Airy beams in order to provide a dynamical
understanding of the autofocusing phenomenon observed with circular Airy beams.
\end{abstract*}

%%%%%%%%%%%%%%%%%%%%%%%%%%%%%%%%%%%%%%%%%%%%%%%%%%%%%%%%%%%%%%%%%%%%%%%%
%%%%%%%%%%%%%%%%%%%%%%%%%%%%%%%%%%%%%%%%%%%%%%%%%%%%%%%%%%%%%%%%%%%%%%%%

\section{Introduction}
\label{sec1}

Airy beams are solutions to the paraxial Helmholtz equation known for exhibiting
shape invariance while the undergo self-acceleration when propagating in free space.
Since their experimental implementation by Christodoulides and coworkers
\cite{christodoulides:PRL:2007}, this peculiar type of structured light has awaken
much interest at present because of its potential applications
\cite{christodoulides:Optica:2019,wong:FrontPhys:2021}.
Nonetheless, although they are receiving much attention within the field of optics, they
were first proposed in the field of quantum mechanics.
As shown by Berry and Balazs \cite{berry:AJP:1979} by the late 1970s, Schr\"odinger's
equation admits a type of non-diffractive solutions with its amplitude having the functional
form of an Airy function (a formal derivation and proof of these solutions was provided
twenty years later by Unnikrishnan and Rau \cite{unnikrishnan:AJP:1996}).
Thus, leaving aside the phase accumulated with time of these solutions, their amplitude is
preserved in time all along the propagation.
Hence, the associated probability density (or the intensity distribution, in the case of
light beams) remains shape invariant.
Nonetheless, it is also seen that these amplitudes undergo a uniform acceleration with time
(they are self-accelerating) without the action of any external potential, which apparently
seems to challenge Ehrenfest's theorem.
Obviously, this is only the price to be paid for keeping the shape invariance and no
fundamental quantum law is violated, as it can readily be noticed when computing the average
position, which turns out to be zero at any time (i.e., the average position does not undergo
any acceleration).

As it was pointed out by Greenberger \cite{greenberger:AJP:1980}, arguing in terms of the
equivalence principle, although there is no external action, this free system can be shown
to be equivalent to a free falling one, which renders ``a more physically transparent
interpretation'', quoting Greenberger himself.
More specifically, he finds that these Airy functions are eigensolutions of the
Schr\"odinger equation describing a particle acted by a uniform gravitational field,
just like plane waves are eigenfunctions when external fields are lacking.
These solutions were actually discussed by Greenberger and Overhauser earlier on
\cite{greenberger:RMP:1979} in the context of the experiment performed by Collela,
Overhauser, and Werner in 1975 (the so-called COW experiment) \cite{colella:PRL:1975}
to determine the action of the uniform acceleration of the Earth's gravitational field on
the phase shift undergone by neutrons in a Laue-type interferometer.
Interestingly, earlier on, considering a different approach, Gibbs showed
\cite{gibbs:AJP:1975} that the eigenvalues of a point-like particle falling under the action
of gravity and bouncing upwards again off a flat surface with no energy losses were give
precisely by the Airy function.
Specifically, the eigenvalues of this so-called quantum bouncer are proportional to the
corresponding positions of the nodes of the Airy function (measured with respect to $x=0$)
\cite{gibbs:AJP:1975,desko:AJP:1983,geabanacloche:AJP:1999}, which has been experimentally
investigated by Nesvizhevsky and coworkers
\cite{nesvizhevsky:NuclInstMethPhys:2000,nesvizhevsky:Nature:2002,nesvizhevsky:EurPhysJC:2005}.

In the case of light, it is clear that interpretations cannot be regarded to the action of
an external gravitational field.
Yet, the idea of recasting and investigating the flow associated with Airy beams in terms of
``bending'' rays is rather appealing.
To this end, it is quite enlightening the alternative picture of light diffraction
summarized in Born's and Wolf's renowned ``Principles of Optics'' \cite{bornwolf-bk},
where the typical wave picture is substituted by a series of wiggling rays bending in
conformity with the wavy behavior displayed by the phase after finding a conducting
half-plane.
These results were published in 1952 by Braunbek and Laukien \cite{braunbek:Optik:1952},
taking as a basis the solutions provided to edge diffraction by Sommerfeld
\cite{sommerfeld:MathAnn:1896} and the flux lines method proposed by Braunbek
\cite{braunbek:ZNaturA:1951} in analogy to the relationship of rays with constant phase
surfaces in geometrical optics.
This method was applied to Young's two slits by Prosser later on \cite{prosser:ijtp:1976-1},
and more recently it has also been generalized to include polarization
\cite{sanz:PhysScrPhoton:2009,sanz:AnnPhysPhoton:2010,sanz:JRLR:2010,sanz:PhysScr:2013}.
These depictions of light showed a good correspondence with the outcomes obtained in the
implementation of Young's experiment with single photons by Steinberg and his group about
a decade ago \cite{kocsis:Science:2011}, where the use of the so-called weak measurements
\cite{aharonov:PRL:1988,sudarshan:PRD:1989} allows the experimental measure of the transverse
momentum distribution under paraxial conditions without suppressing the final Young-type
fringes.

Thus, taking into account the above theoretical framework, here an analysis of the
propagation of Airy beams is presented and discussed in terms of the associated trajectories
or streamlines with the purpose to better understand the transverse energy flow that makes
special this type of structured light, exploring both infinite and finite energy conditions
It is known that the shape invariance and self-acceleration displayed by ideal Airy beams is
connected to the fact that they are not square integrable, that is, that they carry infinite
energy.
Hence, to circumvent this drawback, families of so-called finite-energy Airy-type beams have
been proposed in the literature and, in some cases, also implemented in the laboratory.
In this regard, the trajectory picture here provided is intended to render some clues on the
underlying mechanism that makes them to differ from ideal Airy beams, on the basis of how
energy flows in space rather than on considering a standard analysis of the properties of
the corresponding angular spectrum.
As it is shown, while the foremost part of the beam can be clearly and unambiguously 
associated with the well-known accelerating term, the rear part of the beam corresponds to
a nearly homogeneous distribution of flow trajectories, particularly for large propagation
distances.
This is shown to be related with an effective transfer of trajectories between adjacent lobes
(gradually, from the fore part of the beam to its rear part), which leads to smearing out the
transverse flow along the rear part of the beam.
This is sharp contrast to the situation found in ideal Airy beams, where trajectories
belonging to a given lobe of the intensity distribution remain the same all along the
propagation.
The analysis is supplemented with an also trajectory-based description of Young's
experiment performed with finite-energy Airy beams.
This apparently simple model arises motivated by the autofocusing phenomenon observed with
circular Airy beams \cite{christodoulides:OptLett:2010,christodoulides:OptLett:2011-2}

Accordingly, the work is organized as follows.
The analysis of ideal infinite-energy Airy beams is provided in Sec.~\ref{sec2}, while its
counterpart in the case of finite-energy Airy beams is presented in Sec.~\ref{sec3}.
The analysis of Young-type interference with two counter-propagating Airy beams is presented
and discussed in Sec.~\ref{sec4}.
Finally, a series of concluding remarks are summarized in Sec.~\ref{sec5}.

%%%%%%%%%%%%%%%%%%%%%%%%%%%%%%%%%%%%%%%%%%%%%%%%%%%%%%%%%%%%%%%%%%%%%%%%
%%%%%%%%%%%%%%%%%%%%%%%%%%%%%%%%%%%%%%%%%%%%%%%%%%%%%%%%%%%%%%%%%%%%%%%%

\section{Infinite energy Airy beams}
\label{sec2}

Consider a monochromatic scalar field propagating in free space under paraxial conditions.
As is well known, its behavior along the transverse direction can be parameterized in terms
of the coordinate along which its propagation takes place.
Thus, if propagation mainly takes place along the $z$ direction, then the transverse
behavior can be specified by ${\bf r}_\perp = (x,y)$, on planes perpendicular to the
$z$-axis.
At each $z$ value, the behavior of the field amplitude, henceforth denoted as
$\Psi({\bf r}_\perp,z)$, is described by the paraxial form of Helmholtz's equation,
\be
 i\frac{\partial \Psi({\bf r}_\perp,z)}{\partial z} =
  - \frac{1}{2k} \nabla_\perp^2 \Psi ({\bf r}_\perp,z) ,
 \label{eq1}
\ee
where $\nabla_\perp = (\partial/\partial x, \partial/\partial y)$ and $k = 2\pi n/\lambda_0$,
with $n$ being the refractive index of the medium and $\lambda_0$ the wavelength in vacuum.
Since the aim of this work is the analysis of the effective flux transfer along the
transverse direction (rotations in the transverse plane are disregarded), let us only
consider one transverse coordinate.
Thus, instead of Eq.~(\ref{eq1}), we consider
\be
 i\frac{\partial \psi (x,z)}{\partial z} =
  - \frac{1}{2k} \frac{\partial^2 \psi(x,z)}{\partial x^2} ,
 \label{eq2}
\ee
where $\psi$ is used instead of $\Psi$ to avoid possible misunderstandings.
To further simplify notation, reduced dimensions will also be used in the formal analysis,
specifically, $\tilde{x} = x/x_0$ and $\tilde{z} = z/k x_0^2 = \lambda_0 z /2\pi n x_0^2$, where $x_0$ refers to a transverse distance of interest.
This change of coordinates allows to recast Eq.~(\ref{eq2}) as
\be
 i\frac{\partial \psi(\tilde{x},\tilde{z})}{\partial \tilde{z}} =
  - \frac{1}{2} \frac{\partial^2 \psi(\tilde{x},\tilde{z})}{\partial \tilde{x}^2} .
 \label{eq3}
\ee
Regarding the numerical values for the above introduced parameters, as in
Ref.~\cite{christodoulides:PRL:2007}, here $n = 1$, $\lambda_0 = 488$~nm, and
$x_0 = 53$~$\mu$m.
Accordingly, $\tilde{x} \approx 18.87 x$ and $\tilde{z} \approx 27.65 \times 10^{-3} z$,
with both $x$ and $z$ in mm.

As it was shown by Berry and Balazs \cite{berry:AJP:1979}, the only non-diffractive
(non dispersive) solution to Eq.~(\ref{eq3}) takes the form of an Airy function
\cite{NIST:DLMF},
\be
 \psi(\tilde{x},\tilde{z}) = e^{i(\tilde{x} - \tilde{z}^2/6) \tilde{z}/2}
  Ai(\tilde{x} - \tilde{z}^2/4) ,
 \label{eq4}
\ee
with
\be
 Ai(s) = \frac{1}{\pi} \int_0^\infty \cos \left( \frac{u^2}{3} + su \right) du ,
 \label{eq5}
\ee
which keeps shape invariance all along its propagation at the expense of being a non square
integrable function.
Furthermore, it can also be seen that the beam amplitude (\ref{eq4}) exhibits a quadratic
displacement with the longitudinal coordinate, $z$, along the transverse direction, $x$.
That is, any point of the beam amplitude or the associated intensity can readily be referred
to the input beam,
\be
 \psi(\tilde{x},0) = Ai(\tilde{x}) ,
 \label{eq6}
\ee
by simply adding the quantity $\tilde{x}_d = \tilde{z}^2/4$ or, in terms of the physical
dimensions, $x_d = \lambda_0^2 z^2/16\pi^2 n^2 x_0^3 \approx 10.13 \times 10^{-6}$~mm$^{-1}$.
Interestingly, in spite of showing an evidently accelerated-like displacement, by using
the definition (\ref{eq5}), it can easily be seen that the average position of these beams
vanishes, that is, it seems to be no influence on the average transverse flow of energy.

To better understand this apparently paradoxical aspect, let us determine the transverse
energy flux \cite{bornwolf-bk}, which reads as
\be
 j(\tilde{x},\tilde{z}) = {\rm Im} \left[
  \psi^*(\tilde{x},\tilde{z}) \frac{\partial \psi (\tilde{x},\tilde{z})}{\partial \tilde{x}}
  \right]
  = \frac{\tilde{z}}{2} Ai^2(\tilde{x} - \tilde{z}^2/4) .
 \label{eq7}
\ee
%
%\ba
% j(\tilde{x},\tilde{z}) & = & \frac{1}{2i}
%  \left[
%  \psi^*(\tilde{x},\tilde{z}) \frac{\partial \psi (\tilde{x},\tilde{z})}{\partial \tilde{x}}
%  - \psi(\tilde{x},\tilde{z}) \frac{\partial \psi^* (\tilde{x},\tilde{z})}{\partial \tilde{x}}
%  \right]
% \nonumber \\
% & = &
% \frac{\tilde{z}}{2} Ai^2(\tilde{x} - \tilde{z}^2/4) .
% \label{eq7}
%\ea
%
Taking into account that the flux can always be written in the form of a transport
relation, as $j = v I$, where $I(\tilde{x},\tilde{z}) = |\psi (\tilde{x},\tilde{z})|^2 =
Ai^2(\tilde{x} - \tilde{z}^2/4)$ is the beam intensity distribution and $v$ refers to a
drift or transport effective ``velocity'' field acting on $I$, we have
\be
 v(\tilde{x},\tilde{z}) = \frac{d\tilde{x}}{d\tilde{z}} = \frac{\tilde{z}}{2} .
 \label{eq8}
\ee
Accordingly, by integrating over $\tilde{z}$, we obtain a series of flow trajectories or
streamlines that determine how the intensity, $I$, is transported along the transverse
direction as the beam propagates along the $z$ direction.
These trajectories thus show a quadratic parametric dependence on $\tilde{z}$,
\be
 \tilde{x}(\tilde{z}) = \tilde{x}(0) + \frac{\tilde{z}^2}{4} ,
 \label{eq9}
\ee
in correspondence with the quadratic displacement exhibited by the amplitude (\ref{eq4}).

Physically, Eq.~(\ref{eq9}) means that any trajectory starting within the region delimited
by two nodes of $\psi$ (or, equivalently, two vanishing minima of the corresponding
intensity), will remain confined between them at any subsequent value of $z$.
In other words, there is no effective transfer of energy between adjacent regions of the
beam, as al the flow lines (trajectories) remain parallel one another.
This underlying hydrodynamical description thus explains why the Airy beam remains invariant
all along its propagation, namely, because there is not effective transfer of flow; keeping
the beam accelerated in its ahead (or backward) propagation, like a continuously falling
particle, prevents this dispersion or diffraction \cite{greenberger:AJP:1980}.
To some extent, the infinite-reach of the rear part of the beam somehow ``pushes'' the
foremost one in this increasingly longer and longer transverse displacement.

\begin{figure}[!t]
 \centering
 \includegraphics[height=6.5cm]{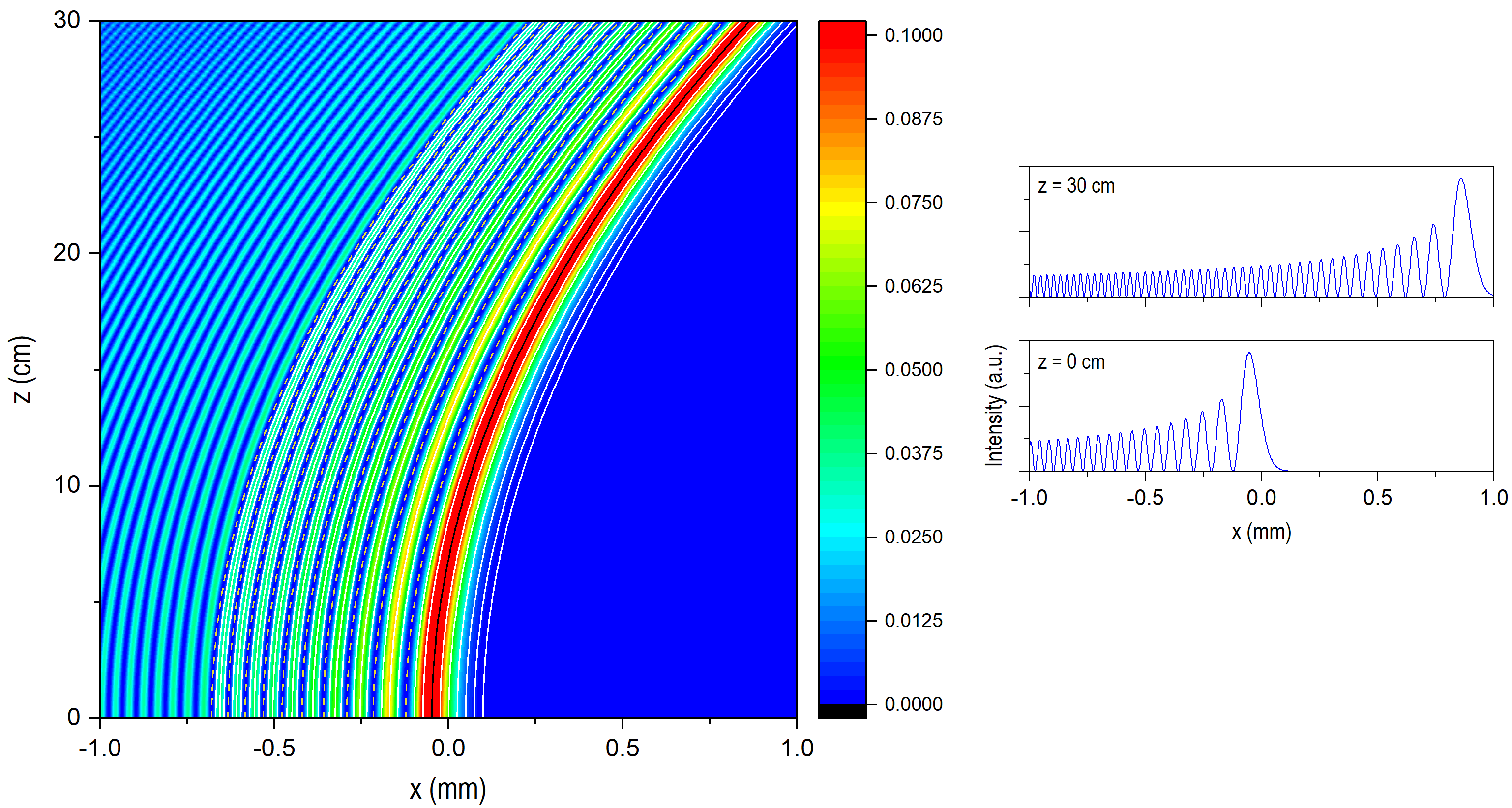}
 \caption{\label{fig1} Propagation of an infinite-energy Airy beam, from $z = 0$~cm to
  $z = 30$~cm.
  The density plot has been truncated at about 78\% its maximum value in order to better
  visualize the evolution of the different secondary maxima and their gradual suppression.
  A set of flux trajectories is superimposed on the density plot in order to illustrate how
  the flux if channeled within adjacent zeros of the Airy function, without mixing with the
  flux associated with other lobes of the intensity distributions.
  To make these jumps more apparent, the evolution of the zeros of the ideal Airy beam has
  been denoted with orange dashed lines.
  The trajectory associated with the center of the principal maximum of the Airy beam is
  represented with a black solid line.
  In the right panels, input ($z = 0$~cm, bottom) and output ($z = 30$~cm, top) intensity
  distributions, normalized to the maximum value to better appreciate the details.}
\end{figure}

This behavior is better understood by inspecting Fig.~\ref{fig1}, where the propagation of
an infinite-energy (ideal) Airy beam, from $z=0$ to $z=30$~cm, is represented in terms of
a density plot.
The plot clearly shows how both nodes and maxima of the intensity distribution follow
parabolic trajectories, making the whole pattern to undergo a net rightwards displacement
of 0.912~mm at $z=30$~cm (this is better seen by comparing the intensity distributions shown
in the two panels on the right side).
When the beam propagation is represented in terms of (flux) trajectories, here indicated
by the set of white solid lines superimposed to the density plot, we note that there is no
mixture of the flux corresponding to different lobes of the intensity distribution.
In other words, the intensity confined within two nodes remains the same all along the
propagation.
To monitor the propagation, we have associated three trajectories to each of the nine
secondary maxima preceding the principal maximum, while nine of them have been associated
to this one, since it is much wider than the other.
Among the nine trajectories associated with the leading lobe, the initial condition of one
of them has been chosen at the position of the maximum (see black solid line).
Although in this particular instance of an ideal Airy beam, this trajectories behaves like
any other, note that under conditions of finite energy (see below) it will serve to detect
how much the beam deviates from ideality and at which $z$ this situation becomes relevant.

%%%%%%%%%%%%%%%%%%%%%%%%%%%%%%%%%%%%%%%%%%%%%%%%%%%%%%%%%%%%%%%%%%%%%%%%
%%%%%%%%%%%%%%%%%%%%%%%%%%%%%%%%%%%%%%%%%%%%%%%%%%%%%%%%%%%%%%%%%%%%%%%%

\section{Finite energy Airy beams}
\label{sec3}

Generating an ideal Airy beam requires an infinite amount of energy as it can readily be
noticed from its flat angular spectrum.
Note that the Fourier transform, $\tilde{\psi}$, of the amplitude (\ref{eq6}) is
\be
 \tilde{\psi}(k_x) =
  \int_{-\infty}^\infty Ai(\tilde{x}) e^{-ik_x\tilde{x}} d\tilde{x} = e^{ik_x^3/3} ,
 \label{eq10}
\ee
which is characterized by a unitary amplitude for all the transverse $k_x$ components.
A way to limit in extension the beam size, thus approaching more realistic working conditions
(both in the laboratory and in numerical simulations), is by gradually suppress the rear
highly-oscillatory tail with a decaying exponential function (i.e., introducing an
exponential-type aperture function), as indicated in \cite{christodoulides:PRL:2007}.
Thus, considering
\be
 \psi(\tilde{x},0) = Ai(\tilde{x}) e^{\gamma \tilde{x}} ,
 \label{eq11}
\ee
with $\gamma$ positive and dimensionless, the propagated solution is given
\cite{christodoulides:PRL:2007,christodoulides:OptLett:2007} by
\ba
 \psi(\tilde{x},\tilde{z}) & = & e^{i(\tilde{x} - \tilde{z}^2/6) \tilde{z}/2
  + \gamma (\tilde{x} - \tilde{z}^2/2) + i \gamma^2 \tilde{z}/2}
  Ai(\tilde{x} - \tilde{z}^2/4 + i \gamma \tilde{z}) \nonumber \\
 & = &  e^{i(\tilde{x} - \tilde{z}^2/6) \tilde{z}/2
  + \gamma (\tilde{y} - i \gamma \tilde{z}/2)} Ai(\tilde{y}) ,
 \label{eq12}
\ea
with
\be
 \tilde{y} = \tilde{x} - \tilde{z}^2/4 + i \gamma \tilde{z} .
 \label{eq13}
\ee
Note that now the solution (\ref{eq12}) represents a finite-energy Airy-type beam,
with its amplitude determined by both a complex Airy function \cite{NIST:DLMF} [now
	$Ai^*(\tilde{y}) = Ai(\tilde{y}^*) \ne Ai(\tilde{y})$ for $\tilde{z} \ne 0$, which
	will eventually lead to a break of the shape-invariance that characterizes real-valued
	Airy-based solutions, as discussed below] and a
modulating prefactor depending on the $\gamma$ parameter.
It is finite in energy, because the exponential term makes it to be a square integrable
function, which can be seen through its angular spectrum \cite{christodoulides:OptLett:2007},
\be
 \tilde{\psi}(k_x) = e^{-\gamma k_x^2 + i(k_x^3 - 3\gamma^2 k_x)/3 + \gamma^3/3} ,
 \label{eq14}
\ee
with a Gaussian amplitude.
And it can be considered an Airy-type beam, because as long as $\gamma$ remains small enough,
the decay of the exponential term in (\ref{eq12}) will be relatively slow and the solution
will be close to an ideal Airy beam, exhibiting both shape invariance and self-acceleration.
Of course, as $\gamma$ increases, the deviation from an ideal Airy beam will be more
evident, with the solution showing a gradual loss of its characteristic highly oscillatory
features displayed beyond the main, leading maximum.
This is a consequence of the asymptotic approach to a Gaussian type distribution, in
compliance with the transverse momentum distribution described by (\ref{eq14}).

Regarding the energy flux, it is clear that the presence of the imaginary component in
the argument of the Airy function is going to play a crucial role, since now the Airy
function becomes complex and hence it is going to actively contribute to $j$.
In particular, after substitution of the wave function (\ref{eq12}) into the first line
of (\ref{eq7}), we obtain
\be
 j(\tilde{x},\tilde{z}) = \left\{
  \frac{1}{2i} \left[ Ai^*(\tilde{y}) \frac{\partial Ai(\tilde{y})}{\partial \tilde{x}}
  - Ai(\tilde{y}) \frac{\partial Ai^*(\tilde{y})}{\partial \tilde{x}} \right]
  + \frac{\tilde{z}}{2} Ai^*(\tilde{y}) Ai(\tilde{y})
 \right\} e^{2\gamma (\tilde{x} - \tilde{z}^2/2)} ,
 \label{eq15}
\ee
which leads to the equation of motion
\ba
 \frac{d\tilde{x}}{d\tilde{z}} & = & \frac{1}{2i} \frac{\partial}{\partial \tilde{x}}
  \left\{ \ln \left[ \frac{Ai(\tilde{y})}{Ai^*(\tilde{y})} \right] \right\}
 + \frac{\tilde{z}}{2} \nonumber \\
 & = & \frac{\partial}{\partial \tilde{x}} \left\{ \arg \left[ Ai(\tilde{y}) \right] \right\}
   + \frac{\tilde{z}}{2} .
 \label{eq16}
\ea
Although Eq.~(\ref{eq16}) cannot be solved analytically, still some physical information of
relevance can be extracted regarding the behavior of the flux.
Comparing this equation of motion with Eq.~(\ref{eq8}), we find that, unless the first term
in Eq.~(\ref{eq16}), on either line, becomes relevant, the behavior of the trajectories in
this latter case is going to be similar to that exhibited by the trajectories associated
with an ideal Airy beam.
Accordingly, it is expected that the trajectories related to the foremost part of the beam
will be less affected (or affected at longer values of $\tilde{z}$) that those in the
rearmost part, where phase variations can be more dramatic.
In the former case, the trajectories will follow a parabola similar to the one described by
Eq.~(\ref{eq9}), at least, until they eventually become affected by finiteness of the beam.
On the other hand, regarding those trajectories in the rearmost part of the beam, note that
the first term in Eq.~(\ref{eq16}) will be of particular relevance whenever the phase
variations become important, which will be close to nodes.
Note that Airy functions have only zeros along the negative real axis; as $\tilde{z}$
increases, though, these zeros move to the right, towards the region of positive $\tilde{x}$.
However, the dependence on the imaginary term $i\gamma\tilde{z}$ is expected to cause a
distortion of the positions of such zeros, thus provoking deviations from the parabolic
motion and making the trajectories to get transferred from one lobe of the intensity to
the adjacent one, just as in the case of Young-type interference
\cite{sanz:AnnPhysPhoton:2010}.

\begin{figure}[!t]
 \centering
 \includegraphics[height=6.5cm]{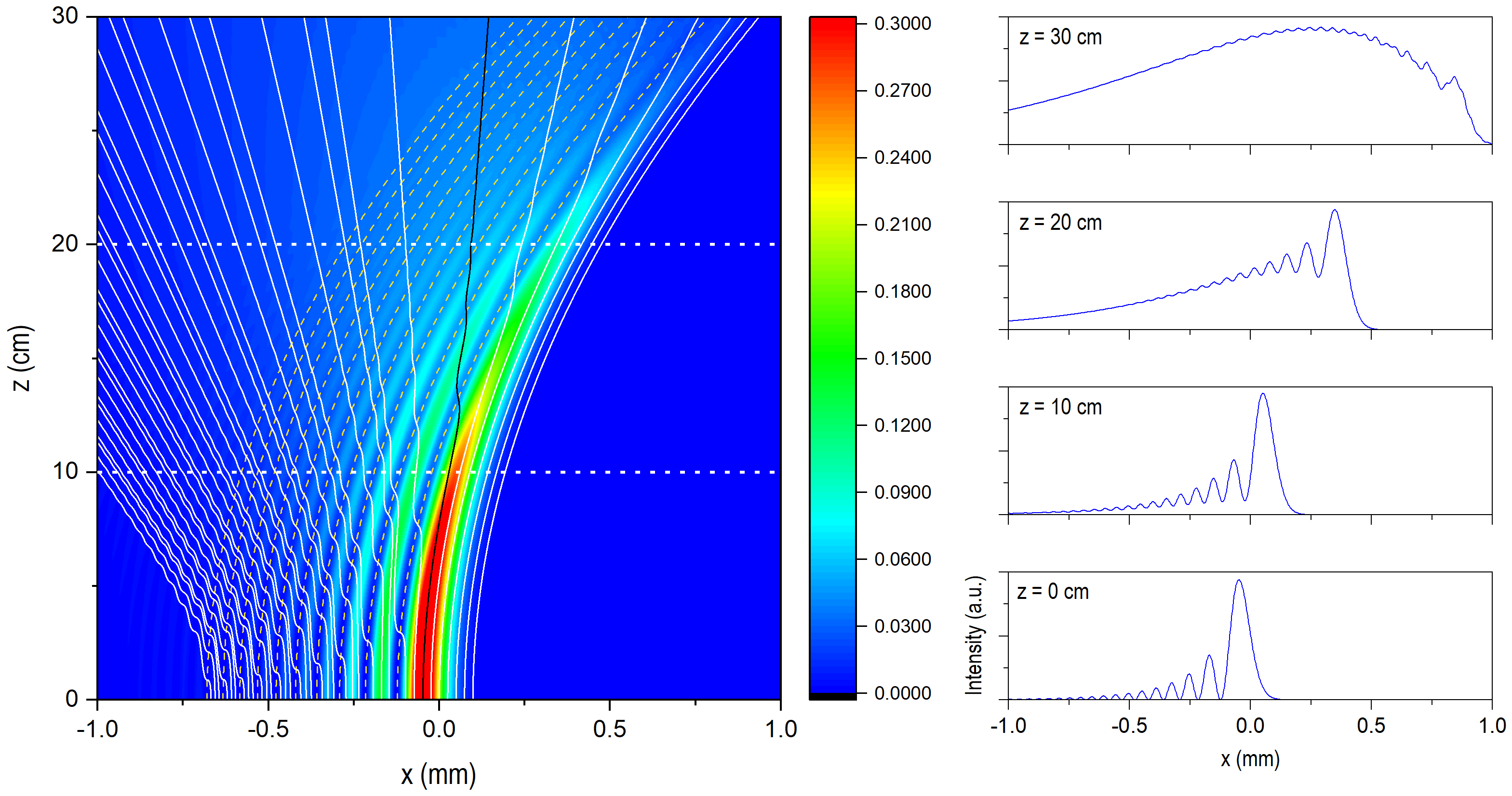}
 \caption{\label{fig2} Propagation of a finite-energy Airy beam with $\gamma = 0.11$, from
  $z = 0$~cm to $z = 30$~cm.
  The density plot has been truncated at about 78\% its maximum value in order to better
  visualize the evolution of the different secondary maxima and their gradual suppression.
  A set of flux trajectories is superimposed on the density plot in order to illustrate how
  the flux jumps gradually from one secondary maximum to the adjacent one, in the direction
  from right to left.
  To make these jumps more apparent, the evolution of the zeros of the ideal Airy beam has
  been denoted with orange dashed lines.
  The trajectory associated with the center of the principal maximum of the Airy beam is
  represented with a black solid line.
  In the right panels, intensity distribution at different values of $z$, normalized to the
  maximum value to better appreciate the details, particularly at large values of $z$, where
  no relevant maxima are observed, but a monotonic trend to a Gaussian distribution centered
  at $x = 0$~mm.
  From bottom to top: $z = 0$~cm, $z = 10$~cm, $z = 20$~cm, and $z = 30$~cm.
  (For an easier identification, the central distributions correspond to the horizontal
  white dotted lines shown in the density plot.)}
\end{figure}

To better understand the above statements, let us consider a numerical simulation based
on the results presented and discussed in
Refs.~\cite{christodoulides:PRL:2007,christodoulides:OptLett:2007}.
The results from such simulation are shown in Fig.~\ref{fig2}, where, as in the case of
Fig.~\ref{fig1}, a density plot of the finite-energy Airy beam has been represented with
a set of trajectories superimposed with the purpose to show how the flux propagates with
$z$.
In this case, both the propagation of the beam and the synthesis of the trajectories have
been numerically computed by means of an ad hoc prepared beam propagation method based on
the split operator technique and an spectral Fourier expansion of the beam \cite{sanz-bk-2}.
Unlike the ideal case, note that here there is an effective transfer of flux in the direction
opposite to the transverse propagation, such that trajectories associated with a given lobe
gradually ``jump'' to the adjacent ones.
Such jumps take place whenever the trajectory crosses a minimum of the intensity distribution
(note that these trajectories cannot cross nodes, but non-vanishing minima), where they
undergo an important boost that allows them to keep propagating in the direction opposite
to the transverse propagation.
Thus, in the rear part of the beam we observe a trend towards a nearly homogeneous spatial
distribution of the trajectories, as the intensity minima are gradually washed out [of 
course, with the proper weight the trajectory distribution should approach a Gaussian
asymptotically, in agreement with the angular momentum distribution (\ref{eq14})].
On the other hand, regarding the leading lobe undergoes a gradual loss of trajectories
with increasing $z$, losing the trajectory associated with its maximum (see black solid
line) at about $z=14$~cm.
This is in correspondence with the intensity distribution profiles displayed in the panels
on the right: while below $z=10$~cm the beam can still be considered of the Airy type, at
$z=20$~cm nearly the main lobe and the adjacent one still remain, and at $z=30$~cm an
incipient asymptotic Gaussian-type distribution already starts becoming apparent.
Nonetheless, the fact that a leading maximum can still be perceived is related to the fact
that a few trajectories (related with a region of low intensity, as it can be noticed from
the position of their corresponding initial conditions) keep track of the self-acceleration
property.

%%%%%%%%%%%%%%%%%%%%%%%%%%%%%%%%%%%%%%%%%%%%%%%%%%%%%%%%%%%%%%%%%%%%%%%%
%%%%%%%%%%%%%%%%%%%%%%%%%%%%%%%%%%%%%%%%%%%%%%%%%%%%%%%%%%%%%%%%%%%%%%%%

\section{Young's interference with finite energy Airy beams}
\label{sec4}

Let us now consider the case of a coherent superposition of two finite Airy beams with the
purpose to analyze the autofucusing in a one-dimensional model, which simplifies the dual
autofocusing associated with circular Airy beams
\cite{christodoulides:OptLett:2010,christodoulides:OptLett:2011-2}
(the model in this work essentially corresponds to a transverse cut of one of such circular
Airy beams).
Given that the two beams are counter-propagating, following the behavior displayed by
analogous two-beam superpositions, it is clear that, at some point, Young-type fringes
should emerge.
Let us thus investigate the issue considering two identical counter-propagating solutions,
each based on a spatial displacement of the input amplitude (\ref{eq11}), namely,
\be
 \psi(\tilde{x},0) = Ai(\tilde{x} - \tilde{x}_0) e^{\gamma (\tilde{x} - \tilde{x}_0)}
  + Ai(\tilde{x} + \tilde{x}_0) e^{\gamma (\tilde{x} + \tilde{x}_0)} ,
 \label{eq17}
\ee
where the value of $\tilde{x}_0 = 0.246$~mm has been chosen to be equal to nearly the width
of the leading lobe of a single Airy beam.

\begin{figure}[!t]
 \centering
 \includegraphics[height=7cm]{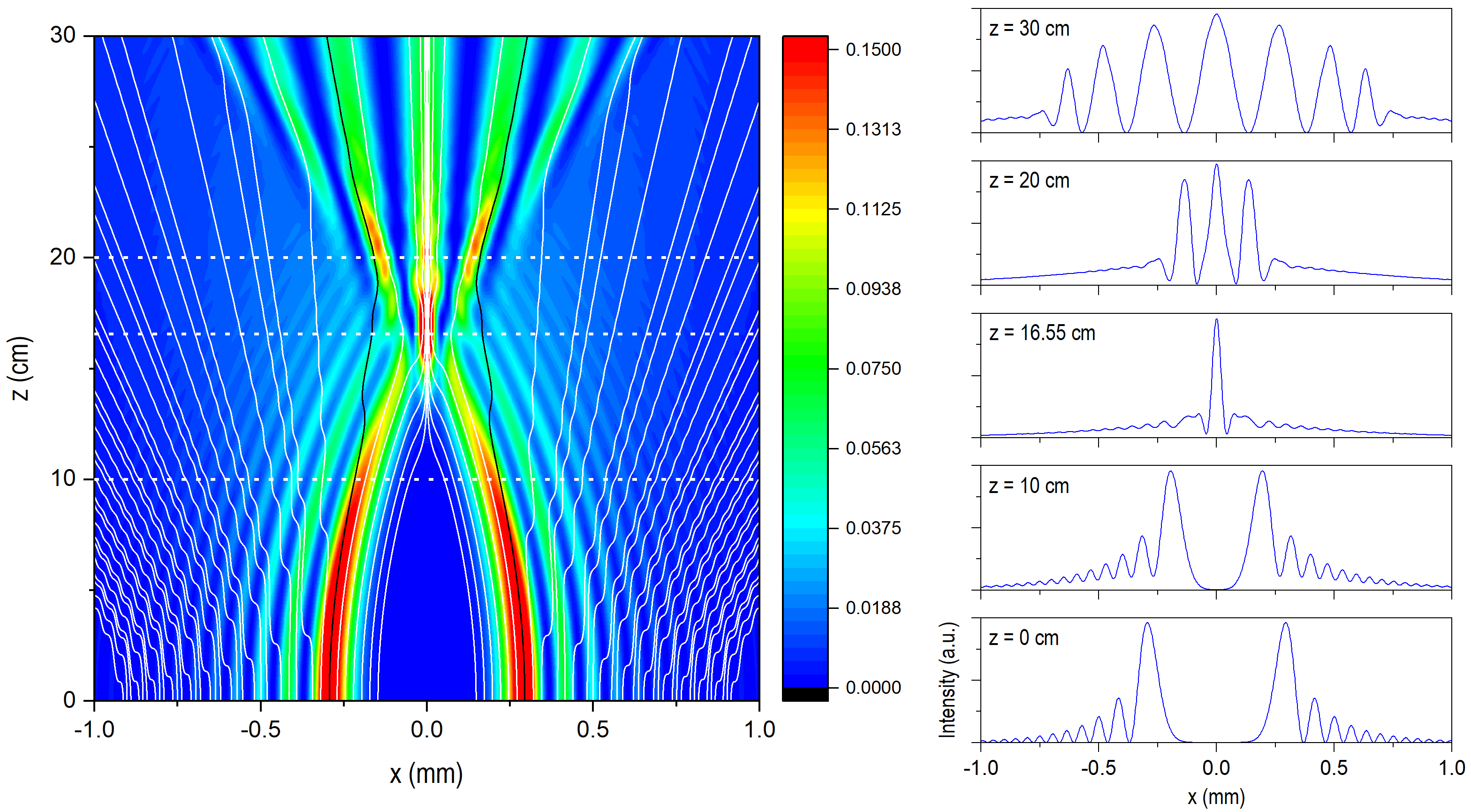}
 \caption{\label{fig3} Propagation, from $z = 0$~cm to $z = 30$~cm, of a superposition of
  two counter-propagating finite-energy Airy beams with $\gamma = 0.11$ and a peak-to-peak distance of about 0.6~mm.
  The density plot has been truncated at about 78\% its maximum value in order to better
  visualize the evolution of the different secondary maxima and the emergence of the
  Young-type interference maxima after the merging of the two Airy beams.
  A set of flux trajectories is superimposed on the density plot in order to illustrate how
  the flux jumps gradually from one secondary maximum to the adjacent one, in the direction
  from right to left.
  Trajectories associated with the center of the principal maximum of each Airy beam are
  denoted with black solid lines.
  In the right panels, the intensity distribution at different values of $z$, normalized to
  the maximum value to better appreciate the details, particularly at large values of $z$,
  where no relevant maxima are observed.
  From bottom to top: $z = 0$~cm, $z = 10$~cm, $z = 20$~cm, and $z = 30$~cm.
  (For an easier identification, the central distributions correspond to the horizontal
  white dotted lines shown in the density plot.)}
\end{figure}

The results from the numerical propagation of the input beam (\ref{eq17}), with
$\gamma = 0.11$, are shown in Fig.~\ref{fig3}, again in the form of a density plot with
sets of trajectories superimposed on the regions covered by both merging beams (the initial
conditions have been chosen as in the two previous cases analyzed, although with the
corresponding spatial displacements on both sides with respect to $x=0$).
To the right, a series of panels show the profile displayed by the intensity distribution as
$z$ increases from the input plane to 30~cm, which stresses the fact that a prominent focused
beam arises along the center of symmetry.
Here, this focusing appears as a strong peak along the $z$ axis, for about 5~cm,
approximately.
However, if we consider a full circular (finite energy) Airy beam instead, the addition of
all contributions arising from the whole circle will lead to an even more prominent on-axis
intensity concentration (with the model used here, such situation would correspond to an
integration over a $\pi$ angle around the $z$-axis).
Beyond this region along $z$, we already observe the emergence of a Young-type fringed
intensity pattern, follow on both sides by a sort of smoothly decaying wings or tails, which
is related to the fact that, far from the input plane, the structure of both beams is washed
out, as seen in Sec.~\ref{sec3}.
Furthermore, also notice that, unlike conventional Young interference, here a gradual
decrease in the width of each lobe is observed, which is related to the fact that the
phase does not depend linearly with $z$.
If, also here, we consider an integration around the $z$ axis, it will be observed a higher
accumulation of intensity along the axis, which is consistent with the autofocusing detected
with circular Airy beams \cite{christodoulides:OptLett:2010,christodoulides:OptLett:2011-2}.
When this behavior is translated in terms of trajectories (see solid white lines), we find
that, if the corresponding trajectories start disseminating until $z=15$~cm in the way
described in Sec.~\ref{sec3}, once the two beams abruptly merge the trajectories start
self-organizing, rejoining within the newly formed intensity lobes.
Moreover, in the merging, the fact that the ``velocity'' field relating the flux with the
intensity is single valued (note that the beam local phase is single valued, except for a
constant factor proportional to $2\pi$) makes the trajectories to avoid their crossing
along the axis, contrarily to what one might expect from usual geometrical rays.
In this regard, because trajectories cannot cross, those corresponding to the centers of
each leading maximum (see solid black lines) will not approach one another, but will bounce
backwards until getting included within the respective second interference orders.

%%%%%%%%%%%%%%%%%%%%%%%%%%%%%%%%%%%%%%%%%%%%%%%%%%%%%%%%%%%%%%%%%%%%%%%%
%%%%%%%%%%%%%%%%%%%%%%%%%%%%%%%%%%%%%%%%%%%%%%%%%%%%%%%%%%%%%%%%%%%%%%%%

\section{Concluding remarks}
\label{sec5}

The use of geometrical rays is widely used to explain the behavior of optical phenomena,
even if they involve a wave treatment of light, as it was formerly considered by Berry and
Balazs \cite{berry:AJP:1979}.
Here, though, it has been shown that trajectories associated with the transverse flux of
energy, in the manner originally proposed by Braunbek \cite{braunbek:Optik:1952} and nicely
summarized by Born and Wolf in their renown monograph \cite{bornwolf-bk}, also constitute a
suitable analysis tool, particularly to understand how the energy flows and hence to elucidate
underlying differences between infinite and finite energy Airy-type beams.
Specifically, it has been shown that, while the foremost part of finite-energy beams can be
clearly and unambiguously associated with the well-known accelerating term, the rear part of
the beam corresponds to a nearly homogeneous distribution of flow trajectories, particularly
for large propagation distances.
This has been shown to be related with an effective transfer of trajectories between adjacent
intensity lobes (gradually, from the fore part of the beam to its rear part), which eventually
leads to progressive a dissemination of the transverse flow along the rear part of the beam.
This is in sharp contrast with the situation found in ideal Airy beams, where trajectories
belonging to a given lobe of the intensity distribution remain the same all along the
propagation.
In sum, we have seen that the trajectories allow us to determine in a more quantitative
manner (more than by simply inspecting the appearance of a density plot) which parts of the
distribution are more importantly affected and how, or which ones are still in compliance
with the self-acceleration that characterize ideal Airy beams.

Furthermore, we have also analyze the interference arising from a coherent superposition of
two counter-propagating finite-energy Airy beams.
Although it is not apparent from a standard density plot, the flux trajectories show a
process of structure loss, denoted by the dissemination of trajectories, which jump from one
intensity lobe to the immediately behind ones, one after the other, until the two leading
maxima of the Airy beams coalesce on the same region along the $z$-axis.
Then, the process reverts and the trajectories start reorganizing again, gathering along
what seems to be a Young-type interference pattern, with an important accumulation of
trajectories that move nearly parallel to the axis, thus explaining the emergence of
autofocusing.

%%%%%%%%%%%%%%%%%%%%%%%%%%%%%%%%%%%%%%%%%%%%%%%%%%%%%%%%%%%%%%%%%%%%%%%%
%%%%%%%%%%%%%%%%%%%%%%%%%%%%%%%%%%%%%%%%%%%%%%%%%%%%%%%%%%%%%%%%%%%%%%%%

%\begin{backmatter}
%\bmsection{Disclosures}
%The author declares no conflicts of interest.

%\bmsection{Data availability}
%All data generated and analyzed during this study are included in the presented research.
%\end{backmatter}

%\begin{backmatter}

%\bmsection{Funding}
\vspace{.15cm}
\noindent {\bf Funding}
This article has no associated award funding.

%\bmsection{Disclosures}
\vspace{.15cm}
\noindent {\bf Disclosures}
The author declares no conflicts of interest.

%\bmsection{Data availability}
\vspace{.15cm}
\noindent {\bf Data availability}
All data generated and analyzed during this study are included in the presented research.

%\end{backmatter}

%%%%%%%%%%%%%%%%%%%%%%% References %%%%%%%%%%%%%%%%%%%%%%%%%

%Add references with BibTeX or manually.

%%%%%%%%%% If using BibTeX:

%\bibliography{references}

%%%%%%%%%% If preparing manually:

\end{document}